\documentclass[twocolumn,amsmath,amssymb, aps]{revtex4}

\usepackage{graphicx}% Include figure files
\usepackage{dcolumn}% Align table columns on decimal point
\usepackage{bm}% bold math
\newcommand{\etal}{\emph{et al.}}
\newcommand{\bs}[1]{\boldsymbol{#1}}
\newcommand{\be}{\begin{equation}}
\newcommand{\ee}{\end{equation}}
\newcommand{\bfig}{\begin{figure}}
\newcommand{\efig}{\end{figure}}
\newcommand{\incl}{\includegraphics}
\begin{document}

\title{Thermal Hall Conductivity as a Probe of Gap Structure in
  Multi-band Superconductors: The Case of $\rm Ba_{1-x}K_xFe_2As_2$}

\author{J. G. Checkelsky$^{1,\dagger}$, R.~Thomale$^2$, Lu Li$^{1,\ddagger}$, G. F. Chen$^3$, J. L. Luo$^3$, N. L. Wang$^3$ and N. P. Ong$^1$
}
\affiliation{
\mbox{$^1$Department of Physics, Princeton University, Princeton, NJ 08544, USA}\\
\mbox{$^2$Department of Physics, Stanford University, Stanford, CA
  94305, USA}\\
\mbox{$^3$Beijing National Laboratory, Institute of Physics, Chinese Academy of Sciences, Beijing 100080, China}
}

\date{\today}
\pacs{}
\begin{abstract}
The sign and profile of the thermal Hall conductivity $\kappa_{xy}$ gives important
insights into the gap structure of multi-band superconductors.
With this perspective, we have investigated $\kappa_{xy}$ and the thermal conductivity $\kappa_{xx}$ in $\rm Ba_{1-x}K_xFe_2As_2$ which
display large peak anomalies in the superconducting state.
The anomalies imply that a large hole-like quasiparticle (qp) population
exists below the critical temperature $T_c$. 
 We show that the qp mean-free-path inferred from $\kappa_{xx}$
reproduces the observed anomaly in $\kappa_{xy}$, providing a consistent estimate of a large qp population.  Further, we demonstrate
 that the hole-like signal is consistent with a theoretical scenario where
despite potentially large gap variations on the electron pockets, the minimal
homogeneous gap of the superconducting phase resides at a hole pocket.  Implications for probing the gap structure in the broader
class of pnictide superconductors are discussed.

\end{abstract}

\maketitle                   % Produces the title
The discovery~\cite{kamihara-08jacs3296,Ren,chen-08prl247002,chen-08n761,Cruz} of superconductivity
in the iron pnictides has galvanized intense interest in
this new class of superconductors.  
As in the cuprates, one of the key issues has been the determination
of the gap symmetry ~\cite{basov-nature}. While for a large class of pnictides theory has quickly converged on an $s_\pm$ order
parameter which changes sign between hole and electron
pockets~\cite{mazin-08prl057003,kuroki-08prl087004,stanev-08prb184509,chubukov-08prb134512,graser-09njp025016}, many
questions remain regarding the actual form of superconducting pairing, such as
gap anisotropies, awaiting further experimental investigation~\cite{hirschi}.  
Among the evidence from measurements on $\rm Ba_{1-x}K_xFe_2As_2$, nuclear magnetic resonance (NMR) relaxation
experiments~\cite{NMR,matano-09epl27012,fukazawa,yashima} appear to be consistent with a multi-gap
scenario of singlet pairing, and penetration depth~\cite{hashimoto-09prl207001} as well as thermal conductivity~\cite{luo-09prb140503}
experiments suggest the presence of strong gap variations inducing
close-to-nodal behavior. Alternatively,
angle-resolved photoemission spectroscopy (ARPES)
experiments ~\cite{ding-08epl47001,wray-08prb184508,ding-11jpc135701}
favor an isotropic multiple gap scenario, with higher confidence for
the hole pockets located at $\Gamma$ than for the electron pockets at
$\text{M}$. 

If the gap parameter $\Delta({\bf k})$ is isotropic on
each FS sheet, the population of
Bogoliubov quasiparticles decreases sharply below $T_c$ ($\bf k$ is a wave vector on the FS).  By contrast, if
nodes exist in $\Delta({\bf k})$ (or if $|\Delta({\bf k})|$
is strongly anisotropic),
the quasiparticle (qp) population decreases quite gradually.
In effective single band superconductors with unconventional gap symmetry,
the thermal Hall conductivity $\kappa_{xy}$ has proved to be a
powerful probe for quasiparticles (qps).
Unlike the diagonal thermal conductivity $\kappa_{xx}$
which is the sum of the electronic term $\kappa_e$ and
the phonon term $\kappa_{ph}$, the off-diagonal term
$\kappa_{xy}$ is purely electronic.
Together, $\kappa_{xx}$ and $\kappa_{xy}$ have
been used to probe extensively the qp density and their
lifetime in the
cuprate $\rm YBa_2Cu_3O_y$ (YBCO)~\cite{Krishana,Zeini,Zhang}
and the heavy fermion superconductor CeCoIn$_5$~\cite{Izawa,Onose}.
In contrast to effective single-band descriptions of the above compounds, the
pnictides are manifestly multi-band superconductors with both electron
and hole-like Fermi pockets. We report thermal Hall results which
connect to the following insight:  {\it whether the
thermal Hall signal is 
electron- or hole-like yields non-trivial information in the
presence of potentially both hole-like and electron-like low-lying
charge carriers.}
In principle, this also applies to thermopower experiments ~\cite{yan}, though such electrical probes
are restricted to the non-superconducting state.
For the pnictides, this provides valuable consistency checks of
different theoretical gap scenarios, as we explicate in the
following for the specific case of $\rm Ba_{1-x}K_xFe_2As_2$. There, we find
theoretically that
while the largest gap anisotropy exist along the electron pockets,
the lowest gap resides at a hole pocket, which is
consistent with our findings from thermal Hall conductivity.

%%%%%%%%%%%%%%%%%%%%%%%%%%%%%%%%%%%%%%%%
%%%%%%%%%%%%%%%%%%%%%%%%%%%%%%%%%%%%%%%%
%%%%%%%%%%%%%%%%%%%%%%%%%%%%%%%%%%%%%%%%
%%%%%%%%%%%%%%%%%%%%%%%%%%%%%%%%%%%%%%%% FIGURE

\bfig[t]            % Fig 1
\incl[width=8.5cm]{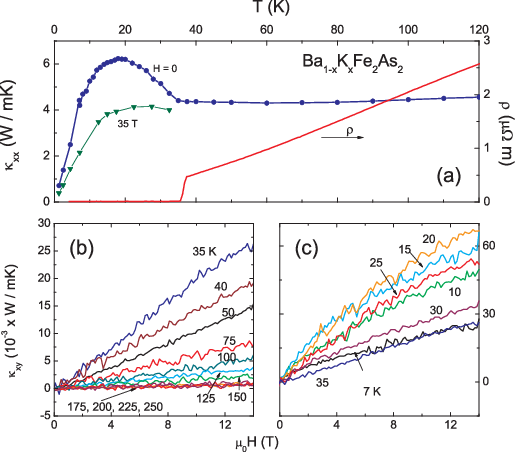}
\caption{\label{Fig1}
(color online) (a) $\kappa_{xx}$ of single-crystal
$\rm Ba_{1-x}K_xFe_2As_2$ at $H$ = 0 (circles)
and at 35 T (triangles).  The solid curve is the
in-plane resistivity $\rho$ in zero $H$ ($T_c$ = 37 K).
$\kappa_{xy}$ vs. $H$ at $T$ = 250$\rightarrow$35 K 
and from 35$\rightarrow$7 K are shown in (b) and (c), respectively. Above $T_c$ (37 K), $\kappa_{xy}$
is $H$ linear up to 14 T, but as $T$ decreases below $T_c$, curvature becomes increasingly apparent.
At all $T$, the Hall signal is hole-like.
}
\efig

To begin, we report detailed measurements of $\kappa_{xx}(T,H)$ and $\kappa_{xy}(T,H)$
on single crystals of $\rm Ba_{1-x}K_xFe_2As_2$ in the geometry
with the field $\bf H||\hat{z}||\hat{c}$ and $-\nabla T||{\bf\hat{x}}$.
The longitudinal and transverse temperature gradients $\delta T_x$
and $\delta T_y$ were measured using chromel-alumel thermocouples.  The 2 crystals studied have dimensions
$2\times 1\times 0.1$ mm$^3$.  At 10 K, the resolution achieved is $\delta T_y\sim 10$ mK.  Measurements of $\kappa_{xy}$ were
made to a field up to 14 T using thermocouples.
To extend measurements of $\kappa_{xx}$ below 6 K
where the thermocouple
sensitivity falls steeply, we used matched
RuO$_x$ micro-sensors
which are very sensitive below 4 K (measurements were performed to $H$ = 35 T).
Electrical measurements are performed using standard 4-probe techniques using a lock-in amplifier.  All measurements are performed in a vacuum atmosphere.

Figure \ref{Fig1}(a) plots the $T$ dependence of $\kappa_{xx}(T,H)$ in Sample 1
with $H=0$ (circles) and with $H$ = 35 T (triangles),
together with the in-plane resistivity $\rho$ (solid curve).
The curve for $\kappa_{xx}(T,0)$ is closely similar in Sample 2.
As shown, $\kappa_{xx}(T,0)$ is nearly $T$ independent between $T_c$ (= 37 K) and
100 K.  Below $T_c$, it rises to a broad maximum that peaks near $\frac12 T_c$.
In a 35-Tesla field, the anomaly is almost completely suppressed.  
In the Boltzmann theory approach, $\kappa_e$
from qp excitations is given by~\cite{BRT,Tewordt}
\be
\kappa_e = \frac{4}{T}\sum_{\bf k}
\left(-\frac{\partial f_0}{\partial E_{\bf k}}\right)
E_{\bf k}^2 v_x({\bf k})^2\tau({\bf k}),
\label{BRT}
\ee
where $E_{\bf k} = \sqrt{\Delta({\bf k})^2 + \epsilon_{\bf k}^2}$ is the
qp energy with $\epsilon_{\bf k}$ the normal-state energy.
Here, ${\bf v(k)}$ is the qp group velocity, $f_0$ the Fermi-Dirac distribution and $\tau({\bf k})$ the transport relaxation time.
A recent treatment of $\kappa_{xy}$
applied to YBCO is given by Durst \etal~\cite{Durst}.
Both
the anomaly profile and its field suppression are similar to features seen
in YBCO and CeCoIn$_5$. In these unconventional superconductors, $\kappa_{xx}$
also rises to a large peak near $\frac12 T_c$, reflecting a
large qp population and a greatly enhanced (zero-field) qp mean-free-path $\ell_0$.
By contrast, $\kappa_{xx}$ decreases roughly linearly
with $(T_c-T)$ below $T_c$ in the clean $s$-wave
superconductors Pb,Hg and Sn~\cite{BRT,Tewordt} (data
on $\kappa_{xy}$ are unavailable).

The curves of $\kappa_{xy}$ vs. $H$ at fixed $T$ are
displayed in Fig.~\ref{Fig1}(b) and ~\ref{Fig1}(c).  Both above and
below $T_c$, the sign of $\kappa_{xy}$ is positive (hole-like).
Above $T_c$, $\kappa_{xy}$ is
strictly linear in $H$ (up to 14 T).  As $T$ decreases
from 100 K to $T_c$,
the slope of $\kappa_{xy}$ vs. $H$ increases gradually until $T_c$, where
it undergoes a sharp increase.
In the normal state, $\kappa_{xy}$ originates from the Lorentz
force acting on the charge carriers (the Hall effect is also
hole-like).
Below $T_c$, the scattering of qps from pinned
vortex lines possesses a right-left asymmetry which leads to a large $\kappa_{xy}$~\cite{Cleary,Krishana,Durst}.  The asymmetry
originates from the circulation of the supercurrent around the
vortex core and the Volovik effect ~\cite{Volovik}.
The qp Hall current initially scales linearly with
the vortex line density $n_V = |B|/\phi_0$, where $B$ is the
flux density and $\phi_0$ the superconducting flux quantum.
At large $B$, however, vortex scattering also reduces
the mean-free-path of the qps (see below),
resulting in a negative curvature in $\kappa_{xy}(H)$.
In Fig.~\ref{Fig1}(c), the steady increase in curvature is
seen as $T$ decreases from 20 to 7 K.

%%%%%%%%%%%%%%%%%%%%%%%%%%%%%%%%%%%%%%%%
%%%%%%%%%%%%%%%%%%%%%%%%%%%%%%%%%%%%%%%%
%%%%%%%%%%%%%%%%%%%%%%%%%%%%%%%%%%%%%%%%
%%%%%%%%%%%%%%%%%%%%%%%%%%%%%%%%%%%%%%%% FIGURE
%
\bfig[t]            % Fig 2
\incl[width=8.65cm]{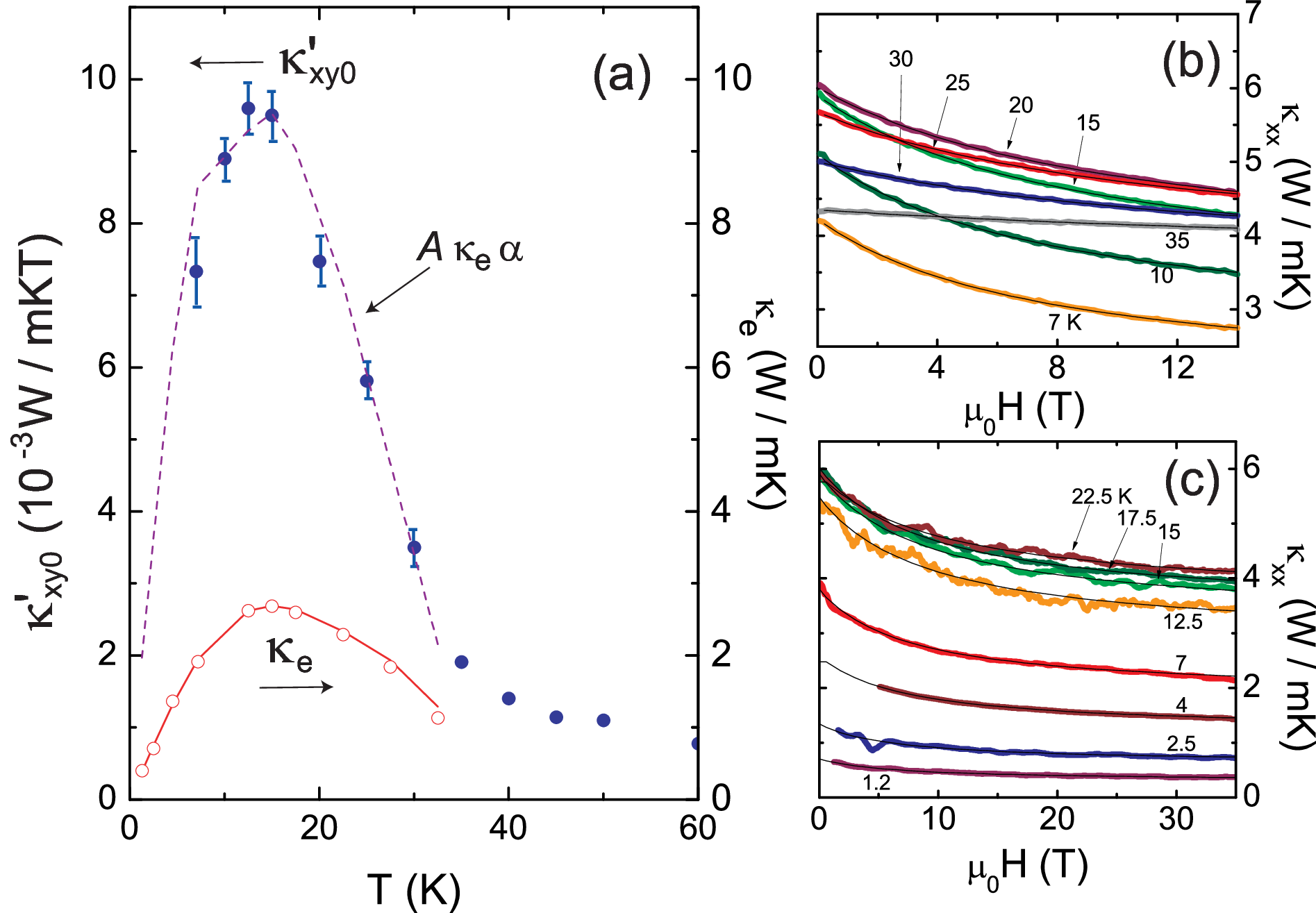}
\caption{\label{Fig2}
(color online) 
(a) $T$ dependence of the weak-field slope
$\kappa'_{xy0} \equiv
\lim_{H\rightarrow 0} \kappa_{xy}/H$ (solid
circles with error bars).  As $T$ falls
below $T_c$, $\kappa'_{xy0}$ displays a steep increase to reach
a peak at 18 K.
For comparison, the quantity $A\alpha\kappa_e$ (with $A$ = 0.033,
see below) is plotted as the dashed curve, and
the electronic term $\kappa_e$, inferred from Eq. \ref{vortex},
is displayed as open circles.
Curves of $\kappa_{xx}$ vs. $H$ at fixed $T$ measured
with thermocouples to 14 T (b) and with RuO$_x$ sensors
to 35 T (c).  In both, fits to Eq. \ref{vortex} are shown as thin curves.
}
\efig

Focusing on the weak-$H$ regime, we see that the sharp
change at $T_c$ in the qp Hall conductivity
$\kappa_{xy}$ is apparent when we
plot the weak-field slope $\kappa'_{xy0} \equiv
\lim_{H\rightarrow 0} \kappa_{xy}/H$
(solid circles in Fig.~\ref{Fig2}(a)).
As $T$ decreases from 200 K to $T_c$,
$\kappa'_{xy}$ increases
slowly by a factor of $\sim$3.  At $T_c$, $\kappa'_{xy0}$
exhibits a sharp break in slope followed
by a steeper rise to a peak that is $\sim$5 times
larger than its value at $T_c$ (curves of $A\kappa_e\alpha$ and
$\kappa_e$ are discussed below).

We next turn to the
diagonal term $\kappa_{xx}(T,H)$, which provides quantitative
estimates of the electronic term $\kappa_e$
and $\ell_0$ independent of $\kappa_{xy}$ (Fig.~\ref{Fig2}(b) and ~\ref{Fig2}(c)).
In the normal state, over the interval 250 to 40 K,
the field dependence of $\kappa_{xx}$ is undetectable with
our sensitivity.  Just below $T_c$, a weak $H$ dependence becomes
apparent (curve at 35 K). As $T$ decreases,
this rapidly evolves to a singular $|B|$ dependence that characterizes the
scattering of qps from pinned vortices in type-II superconductors in the
clean limit ($\ell_0\gg\xi$, where $\xi$ is the coherence length).

The data in Fig. ~\ref{Fig2}(b) and ~\ref{Fig2}(c) fit well to
the vortex-scattering expression (shown as thin curves)~\cite{Cleary,Krishana,Durst}
\be
\kappa_{xx}(T,B) = \frac{\kappa_e(T)}{1+\alpha(T) |B|} + \kappa_{ph}(T),
\label{vortex}
\ee
where $\alpha(T)= \ell_0\sigma_{tr}/\phi_0$ with $\sigma_{tr}$ the
vortex cross-section presented to an incident qp.
Equation \ref{vortex} expresses the
additivity of the zero-$H$ scattering rate ($\sim \ell_0^{-1}$) and  the scattering rate introduced by vortices $\ell_v^{-1}=\sigma_{tr}n_V$. Because $H_{c2}>$80 T,
the condensate amplitude is nearly unaffected by the applied $H$.
Hence we can assume that $\kappa_e(T)$ is nearly
independent of $H$,
and the dominant contribution to the observed $H$ dependence arises
from vortex scattering.  We have assumed that the phonon term $\kappa_{ph}$ has negligible $H$ dependence,
consistent with $q\xi\ll 1$, where
$q$ is the average phonon wave vector for $T<T_c$.

%%%%%%%%%%%%%%%%%%%%%%%%%%%%%%%%%%%%%%%%
%%%%%%%%%%%%%%%%%%%%%%%%%%%%%%%%%%%%%%%%
%%%%%%%%%%%%%%%%%%%%%%%%%%%%%%%%%%%%%%%%
%%%%%%%%%%%%%%%%%%%%%%%%%%%%%%%%%%%%%%%% FIGURE
%
\bfig[t]            % Fig 3
\incl[width=8.5cm]{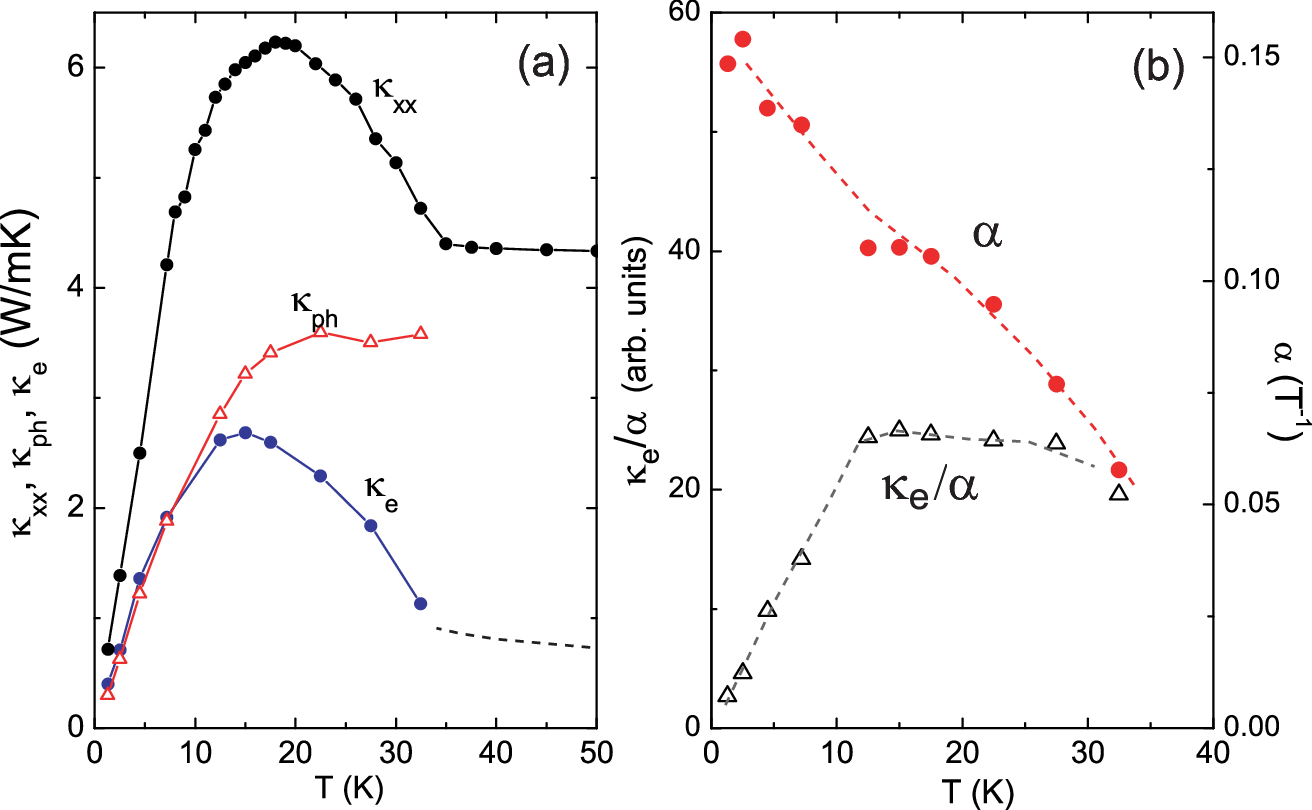}
\caption{\label{Fig3}
(color online) 
(a) Comparison of $\kappa_{xx}$ with $\kappa_e$ and $\kappa_{ph}$
inferred from fits of curves in Fig.~\ref{Fig2}(b) and ~\ref{Fig2}(c) to Eq. \ref{vortex}.
Above $T_c$, $\kappa_e$ is less than $\frac15\kappa_{xx}$.  Below $T_c$,
however, $\kappa_e$ increases rapidly to account for the entire
anomaly in the observed $\kappa_{xx}$. Panel (b) displays
$\alpha = \ell_0\sigma_{tr}/\phi_0$ (solid circles) and $\kappa_e/\alpha$ (open triangles) obtained from the fits.  With $\sigma_{tr}\sim$ 26 \AA,
we estimate that $\ell_0\simeq$ 1,200 \AA~at 2 K.  The quantity $\kappa_e/\alpha$ is a measure of the qp population below $T_c$.}
\efig

From the fits to the curves in Fig.~\ref{Fig2}, we
have determined $\kappa_e(T)$, $\kappa_{ph}(T)$
and $\alpha(T)$.  As shown in Fig. \ref{Fig3}(a), the phonon term $\kappa_{ph}$
(open triangles) decreases monotonically as $T$ decreases below $T_c$.
By contrast, the zero-$H$ electronic term $\kappa_e$
rises to a broad maximum at 15 K and then falls.
As evident, the monotonic profile of $\kappa_{ph}$ implies
that the peak in $\kappa_{xx}$ (in zero $H$)
is entirely associated with $\kappa_e$.  In turn, the
large peak in $\kappa_e$ demands
a large qp population that survives to $\sim\frac12 T_c$.

The fits also yield estimates of $\alpha =\ell_0\sigma_{tr}/\phi_0$ which we display in Fig.~\ref{Fig3}(b).
As $T$ decreases from 35 to 1.2 K,
$\alpha$ rises to $3\times$ the value at $T_c$.  This
reflects primarily the increase in $\ell_0$.
According to Cleary~\cite{Cleary}, the cross-section $\sigma_{tr}\simeq\xi$.  With the estimate $\xi\simeq$ 20 \AA~(from $H_{c2}\sim$85 T), we find
that $\alpha$ at 1.2 K corresponds to $\ell_0\simeq$ 1,200 \AA.
This supports our assumption that $\rm Ba_{1-x}K_xFe_2As_2$ is in the
clean limit $\ell_0\gg\xi$ as has been widely reported.
Since $\kappa_e$ is proportional to $\ell_0$, a nominal
picture of the $T$ dependence of the qp population may be obtained
from the ratio $\kappa_e/\ell_0$.  This quantity is plotted as
open triangles in Fig.~\ref{Fig3}(b).  Initially,
$\kappa_e/\ell_0$ is nominally $T$-independent,
but decreases linearly with $T$ below 13 K. 

The fits of $\kappa_{xx}$ vs. $H$ to Eq. \ref{vortex} has allowed us to
determine $\kappa_e$, $\kappa_{ph}$, and $\alpha$ independent
of $\kappa_{xy}$.
To show that these estimates are consistent with the thermal
Hall conductivity, we note that $\kappa_{xy0}'\sim\ell_0^2$ should
share the same $T$ dependence as $\alpha\kappa_e\sim\ell_0^2$, in the
semi-classical approximation.  In Fig.~\ref{Fig2}(a),
we have plotted $A\kappa_e\alpha$ with the scale-factor $A= 0.033$ (dashed curve) to compare it with the measured values
of $\kappa_{xy0}'$.  Within the uncertainty inherent in
$\kappa_{xy0}'$, the $T$ dependences may be seen to track each other
quite well, especially between 12 and 35 K.
Hence, both $\kappa_{xx}$ and $\kappa_{xy}$ indicate a large hole-like qp population that
persists down to $T\sim\frac12 T_c$.  Moreover,
the values of $\alpha\sim\ell_0$ inferred from
$\kappa_{xx}$ give a consistent description
of the $T$ dependences of both $\kappa_e$ and $\kappa_{xy0}'$.

%%%%%%%%%%%%%%%%%%%%%%%%%%%%%%%%%%%%%%%%
%%%%%%%%%%%%%%%%%%%%%%%%%%%%%%%%%%%%%%%%
%%%%%%%%%%%%%%%%%%%%%%%%%%%%%%%%%%%%%%%%
%%%%%%%%%%%%%%%%%%%%%%%%%%%%%%%%%%%%%%%% FIGURE 4
\bfig[t]            % Fig 4
\incl[width=8.5cm]{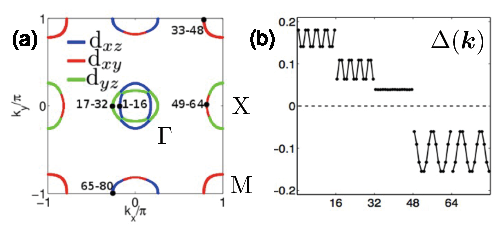}
\caption{\label{Fig4}
(color online)
Band structure (a) and SC gap function (b) for the
    122 band structure around optimally hole-doped filling. The different
    dominant $d$ orbital weights are plotted along the Fermi surface
    in (a)
    indicated by blue, green, and red. The 
    pockets are divided into 80 momentum patches enumerated
    counterclockwise (a). }
\efig

The experimental evidence of large low-lying hole-like qp weight we
obtain from our thermal Hall measurement can be reconciled with a
theoretical perspective on the problem. Starting from a simplified
band structure fit for $\rm Ba_{1-x}K_xFe_2As_2$~\cite{graser-10prb214503}, the Fermi surface topology is
schematically depicted in Fig.~\ref{Fig4}(a) in the unfolded Brillouin
zone with 1 Fe atom per unit cell. As seen there, the $d_{xz}$ and
$d_{yz}$ Fe orbitals dominate the hole pockets at $\Gamma$, while the
third hole pocket at $\text{M}$ is composed of $d_{xy}$. (The third hole
pocket at $\text{M}$ which also maps onto the $\Gamma$ point in the folded Brillouin zone
has been detected in ARPES being of nearly
identical size and shape as one of the other hole
pockets~\cite{ding-11jpc135701,ronnyandrew}.) The electron pockets at
$\text{X}=(\pi,0)/(0,\pi)$ involve weights of all these three $d$
orbitals. Assuming that the inter-pocket scattering is dominated by intra-orbital interactions, various
theoretical approaches predict a significant gap anisotropy on the
electron pockets as a consequence of frustrated electron-electron and
electron-hole scattering~\cite{thomale-11prl187003,platt-11ap638},
while the hole pocket gaps are assumed rather
homogeneous. In a 4 pocket scenario where the hole pocket at $\text{M}$ is
absent, it would hence be a natural guess that the weakest gap can be found on the electron pockets~\cite{thomale-11prl187003}, and as a consequence
that the thermal Hall signal in the SC phase should be
electron-like. In contrast, the hole pocket at $\text{M}$ significantly
changes the situation: Fig.~\ref{Fig4}(b) shows the gap function
$\Delta(\bs{k})$ which we have computed by
multi-orbital functional renormalization
group~\cite{wang-09prl047005,thomale-09prb180505,platt-09njp055058,thomale-11prl187003},
with interaction parameters as chosen in~\cite{thomale-11prl117001}. To
begin with, we find the sign change from hole to electron pockets, as being
characteristic for an $s_\pm$ order parameter. The largest gap
anisotropy is found on
the electron pockets. However, we find the
smallest amplitude to be located on the hole pocket at $\text{M}$. This is
because most of the scattering of this pocket is governed
by the subdominant inter-orbital repulsion scale, as only a small part of the electron
pockets share the $d_{xy}$ orbital content. In addition, the
available phase space for qps is quite large on the pocket (or at
least comparable to the other pockets), which is consistent with the
dominant hole-like qp profile of the thermal Hall conductivity measurements.

We propose extending thermal Hall measurements beyond optimally doped $\rm Ba_{1-x}K_xFe_2As_2$ will 
provide an incisive test to gain experimental insight into the nature
of multi-band superconducting pairing mechanisms in general. For example, for further K
doping we expect the hole-like profile to persist and become even more
pronounced, as $d$-wave order can form and even give rise to nodes on the
hole pockets~\cite{thomale-11prl187003}. Moving to the
electron-doped side through Co doping should eventually
remove the broad hole band giving rise to the hole pocket at $\text{M}$ from
the Fermi surface, by which the small gap regimes on the electron
pockets should become the dominant contribution to low-energy charge
carriers. We hence predict a sign change of the thermal Hall signal
as a function of doping. Similar trends may be triggered by stronger
nodal propensity due to isovalent doping of the As-based compound by P, potentially
giving rise to accidental nodes on the electron pockets~\cite{thomale-11prl187003,kuroki}.
It may likewise be interesting to investigate the recently discovered
iron chalcogenides such as
K$_x$Fe$_2$Se$_2$~\cite{guo-10prb180520,fang-selen}, where the
potential role of hole-like carriers may be intimately linked to the
competition between a possible $s_\pm$ and gapped $d$ wave order parameter~\cite{wang-11epl57003,maier-11prb100515,thomale-11prx011009}. 
As a result, it is likely that the properties of the quasiparticles
extracted from heat transport will be valuable for understanding
the pairing mechanism of multi-band superconductors.

We thank P. A. Lee, D.-H. Lee, S. Graser, D. Scalapino, B. A. Bernevig, and M. Z. Hasan
for helpful discussions.
The research at Princeton is supported by U.S. National Science
Foundation (NSF) under Grant DMR-0819860.
Research at IOP is supported by
NSFC, 973 project of MOST
and CAS of China. High-field experiments were performed at the
National High Magnetic Field Laboratory, Tallahassee, a national
facility supported by NSF, the Dept. of Energy and the State of Florida.
RT is supported by an SITP fellowship by Stanford University.

\noindent
\emph{$^\dagger$Present address of JGC}: Advanced Science Institute, RIKEN, Saitama, Japan.
\emph{$^\ddagger$Present address of LL}: Dept. of Physics, University of Michigan, Ann Arbor, MI, USA.

\end{document}